\documentclass[aps,prb,twocolumn,superscriptaddress,amsmath,amssymb]{revtex4}
\usepackage{graphicx}
\usepackage{epstopdf}
\usepackage{longtable}
\usepackage{hyperref}

\usepackage{bm}

\begin{document}

\title{Electronic structure of Fe and magnetism in the $3d/5d$ double perovskites Ca$_2$FeReO$_6$ and Ba$_2$FeReO$_6$}

\author{E. Granado}
\affiliation{``Gleb Wataghin'' Institute of Physics, University of Campinas - UNICAMP, Campinas, S\~ao Paulo 13083-859, Brazil}

\author{J. C. Cezar}
\affiliation{Laborat\'{o}rio Nacional de Luz S\'{i}ncrotron, Caixa Postal 6192, Campinas, S\~ao Paulo 13083-970, Brazil}

\author{C. Azimonte}
\affiliation{``Gleb Wataghin'' Institute of Physics, University of Campinas - UNICAMP, Campinas, S\~ao Paulo 13083-859, Brazil}

\author{J. Gopalakrishnan}
\affiliation{Solid State and Structural Chemistry Unit, Indian Institute of Science,
Bangalore 560012, India}

\author{K. Ramesha}
\affiliation{Solid State and Structural Chemistry Unit, Indian Institute of Science,
Bangalore 560012, India}

\begin{abstract}

The Fe electronic structure and magnetism in (i) monoclinic Ca$_2$FeReO$_6$ with a metal-insulator transition at $T_{MI} \sim 140$ K and (ii) quasi-cubic half-metallic Ba$_2$FeReO$_6$  ceramic double perovskites are probed by soft x-ray absorption spectroscopy (XAS) and magnetic circular dichroism (XMCD). These materials show distinct Fe $L_{2,3}$ XAS and XMCD spectra, which are primarily associated with their different average Fe oxidation states (close to Fe$^{3+}$ for Ca$_2$FeReO$_6$ and intermediate between Fe$^{2+}$ and Fe$^{3+}$ for Ba$_2$FeReO$_6$) despite being related by an isoelectronic (Ca$^{2+}$/Ba$^{2+}$) substitution. For Ca$_2$FeReO$_6$, the powder-averaged Fe spin moment along the field direction ($B = 5$ T), as probed by the XMCD experiment, is strongly reduced in comparison with the spontaneous Fe moment previously obtained by neutron diffraction, consistent with a scenario where the magnetic moments are constrained to remain within an easy plane. For $B=1$ T, the unsaturated XMCD signal is reduced below $T_{MI}$ consistent with a magnetic transition to an easy-axis state that further reduces the powder-averaged magnetization in the field direction. For Ba$_2$FeReO$_6$, the field-aligned Fe spins are larger than for Ca$_2$FeReO$_6$ ($B=5$ T) and the temperature dependence of the Fe magnetic moment is consistent with the magnetic ordering transition at $T_C^{Ba} = 305$ K. Our results illustrate the dramatic influence of the specific spin-orbital configuration of Re $5d$ electrons on the Fe $3d$ local magnetism of these Fe/Re double perovskites.

\end{abstract}

\maketitle

\section{Introduction}

In recent years, considerable attention has been given to $4d$ or $5d$-based materials showing a combination of sizable spin-orbit coupling and electronic correlations. In particular, the spin-orbit entangled $J_{eff}=1/2$ state found in Sr$_2$IrO$_4$ and related iridates is a direct consequence of such combination \onlinecite{Kim,Kim2,Jackeli,Caorev}, with profound physical consequences such as a remarkable similarity between the iridate and cuprate phase diagrams \onlinecite{Kim_ARPES1,Yan,Kim_ARPES2,Samanta}. This new front of research in condensed matter physics is not expected to remain restricted to iridates, but is most likely extensible to other compounds containing magnetic $4d$ or $5d$ ions.
Also, alternating $3d$ and $4d/5d$ ions in an ordered double perovskite structure offers a possible pathway to investigate ground states arising from the combination of strong SOC in $4d/5d$ ions and strong electronic correlation in $3d$ ions \onlinecite{Cavichini}. An interesting and relatively well studied family of $3d/5d$ double perovskites is the $A_2$FeReO$_6$ system ($A$= Ca, Sr, Ba). The compounds Ba$_2$FeReO$_6$ and Sr$_2$FeReO$_6$ are ferrimagnetic half-metals with paramagnetic transition temperatures $T_C$'s above room temperature, being materials of interest for spintronics that are also related to the physics of Sr$_2$FeMoO$_6$ double perovskites (for reviews, see Refs. \onlinecite{Serratereview,DeTeresareview,Vasalareview}). Ba$_2$FeReO$_6$ in particular has a cubic crystal structure above $T_C^{Ba}=305$ K and a slight tetragonal distortion below $T_C^{Ba}$ due to a significant orbital polarization of the Re $5d$ electrons \onlinecite{Ferreira,Azimonte1}. The interplanar Bragg distances follow the direction of an external magnetic field below $T_C^{Ba}$, advocating for a decisive role of a sizable spin-orbit coupling to the physics of this and related materials \onlinecite{Azimonte1}. In fact, a number of x-ray magnetic circular dichroism (XMCD) studies in this family yielded large values of the Re orbital/spin magnetization ratio ($\vline m_{L}/m_S \vline \sim 0.28-0.67$ \onlinecite{Azimonte1,Azimonte2,Sikora1,Sikora2,Escanhoela}). 

The half-metallic and ferrimagnetic ground state of many double perovskites such as Sr$_2$FeMoO$_6$, Sr$_2$FeReO$_6$ and Ba$_2$FeReO$_6$ is normally modelled in terms of double exchange interactions in the presence of spin-polarized conduction electrons \onlinecite{Serratereview,DeTeresareview,Vasalareview}. In fact, the electronic structure of these materials is such that the spin-up Fe $3d$ shell is completely filled, and the electronic band crossing the Fermi level $E_F$ is composed of hybridized Fe($3d$:$t_{2g}$)-O($2p$)-Re($5d$:$t_{2g}$) spin-down levels. Thus, the conduction electrons can be shared by Fe and Re ions in the ferrimagnetic configuration, leading, in an ionic picture, to mixed-valent Fe$^{2+} /$Fe$^{3+}$ and Re$^{6+} /$Re$^{5+}$ ions for Ba$_2$FeReO$_6$ and Sr$_2$FeReO$_6$ \onlinecite{Gopal}.

The double exchange mechanism provides a clear connection between the physics of double perovskites and that of doped manganites. A  distinction between these systems, however, is that strong correlation effects between Mn $3d:e_g$ electrons lead to competing charge and orbitally ordered phases for manganites \onlinecite{manganitereview}, being a decisive factor to unlock a plethora of fascinating physical phenomena and phase transitions that are characteristic of this system. On the other hand, in double perovskites the effects of electronic correlations are somewhat less explored. A candidate for showing strong correlation effects is Ca$_2$FeReO$_6$, the physical properties of which differ strongly from those of Ba$_2$FeReO$_6$ and Sr$_2$FeReO$_6$. For instance, Ca$_2$FeReO$_6$ shows a metal-insulator transition at $T_{MI} \sim 140-150$ K \onlinecite{Kato}, a much lower temperature than the ferrimagnetic ordering temperature of this material, $T_C^{Ca} \sim 520-540$ K \onlinecite{Westerburg,Granado,Alamelu,DeTeresa}. Structurally, the competing phases are characterized by the same monoclinic $P2_1/n$ space group with an $a^{-}a^{-}b^{+}$ tilt pattern in Glazer's notation
but with slightly different lattice parameters \onlinecite{Westerburg,Granado,Kato,Oikawa}. Band structure calculations do not capture the insulating state of Ca$_2$FeReO$_6$ unless a relevant on-site Coulomb repulsion term is introduced \onlinecite{Wu,Jeng,Szotek,Iwasawa,Antonov,Lee,Jeon}, leading to the perception that the Re $5d$ electrons are indeed significantly correlated in this double perovskite \onlinecite{Granado,Iwasawa,Antonov,Lee}. In fact, this transition at $T_{MI}$ has been associated with an orbital ordering transition of the Re $5d$ electrons \onlinecite{Oikawa,Lee,Yuan}. The possible orbital character of the transition $T_{MI}$ does not necessarily imply that the state above $T_{MI}$ is orbitally disordered, nor that the spin-orbital coupling energy is smaller than the exchange coupling in this system. Actually, the monoclinic $b$ lattice parameter of Ca$_2$FeReO$_6$ shows an anomalous expansion below $T_C^{Ca}$, \onlinecite{Granado}, suggesting that an orbitally ordered state associated with the Re spin-orbit coupling develops just below the magnetic ordering transition (i.e., much above $T_{MI}$). This conclusion is also supported by the large magnetostriction of this compound even for $T > T_{MI}$ \onlinecite{Serrate}. Thus, the possible orbital transition at $T_{MI}$ for Ca$_2$FeReO$_6$ most likely involves two distinct and competing spin-orbital ordered states rather than being a transition between an orbitally ordered and a disordered state. According to neutron diffraction experiments \onlinecite{Granado,Oikawa}, the spin-orbital state below $T_{MI}$ shows a spontaneous moment direction along the monoclinic principal axis ({\bf b}), while above $T_{MI}$ the magnetic moments lie in the {\it ac}-plane.
More recently, an inelastic neutron scattering study showed the development of a gap in the spin excitations below $T_{MI}$ \onlinecite{Yuan}, consistent with the development of an easy-axis magnetic state below this temperature. On the other hand, the gapless magnetic excitations above $T_{MI}$ are consistent with either an easy-plane spin-orbital configuration or an orbitally disordered state. Application of a magnetic field of several tesla tends to favor the growth of the metallic state \onlinecite{Granado}. This effect leads to a remarkable magnetoresistance effect for Ca$_2$FeReO$_6$ and Ca$_{1.5}$Sr$_{0.5}$FeReO$_6$ \onlinecite{magnetoresistance}


Despite the considerable attention paid so far to the behavior of the Re $5d$ electrons and the relative role of their electronic correlations and spin-orbit coupling in these Fe/Re-based double perovskites, relatively little information on the element-specific Fe electronic states is presently available. Considering that the valence/conduction electrons in these systems may show a mixed character between Fe $3d$, Re $5d$ and O $2p$ levels, a systematic investigation of the Fe $3d$-projected states may provide additional information about the intriguing physics presented by this system. In this work, we report the Fe electronic structure and magnetism of Ca$_2$FeReO$_6$ and Ba$_2$FeReO$_6$ by means of XAS and XMCD experiments at the Fe $L_{2,3}$ edges, for temperatures below 350 K and magnetic fields below 5 T. The XAS spectra show substantial differences between these compounds, consistent with their distinct Fe $3d$ electronic occupations. The temperature-dependencies of their XAS spectra show anomalies associated with charge-transfer effects at $T_{MI} = 140$ K for Ca$_2$FeReO$_6$ and $T_C^{Ba} = 305 $ K for Ba$_2$FeReO$_6$. In addition, the Fe moment at low temperatures obtained from our XMCD data for Ca$_2$FeReO$_6$ for a field of 5 T is substantially smaller than the Fe moment obtained by a previous neutron diffraction study on the same sample \onlinecite{Granado}, signaling a very strong magnetostructural coupling for the Fe $3d$ moments. For $B=1$ T, the unsaturated XMCD signal is reduced below $T_{MI}$ indicating a magnetically harder insulating state with respect to the metallic one. It is inferred that the observed hardness of the Fe moments in CFRO, and to a lesser extend in BFRO, is caused by a superexchange interaction with the Re $5d$ moments, where the latter are pinned to the lattice due to the sizable spin-orbit interaction. This also leads to the observed sensitivity of the field-aligned Fe $3d$ ordered moments probed by XMCD to the transition at $T_{MI}$, where the specific spin-orbital ordering pattern of the Re $5d$ electrons is most likely changed.

\section{Experimental details}

The pellets of polycrystalline Ca$_2$FeReO$_6$ and Ba$_2$FeReO$_6$ used here are the same employed in our previous investigations \onlinecite{Granado,Azimonte1,Azimonte2}. Details of the synthesis method can be found elsewhere \onlinecite{Prellier}.
The XAS and XMCD measurements were done at the European Synchrotron Radiation Facility (ESRF), on the high field magnet end station of beamline ID08 by taking the total electron yield. The pellets were scrapped prior to the measurements with a diamond file under high vacuum. All XAS [$(\mu_{+}+\mu_{-})$] and XMCD [$(\mu_{+}-\mu_{-})$] experiments were performed with 100 \%\ circularly polarized light under applied magnetic fields along the direction of beam propagation. $\mu_{+}$ and $\mu_{-}$ spectra as a function of $T$ were collected on warming by changing the helicity of the incoming photons at a fixed field, which was applied at the base temperature after a zero field cooling procedure from 300 K. All XAS data were normalized by the edge step between 700 and 730 eV. The spectra taken in different conditions were aligned in energy using a weak feature observed in the incident flux ($I_0(E)$) near the Fe $L_3$ edge position, which was associated with a small Fe contamination in the beamline optics.

\section{Results and analysis}

\begin{figure}
\includegraphics[width=0.45\textwidth]{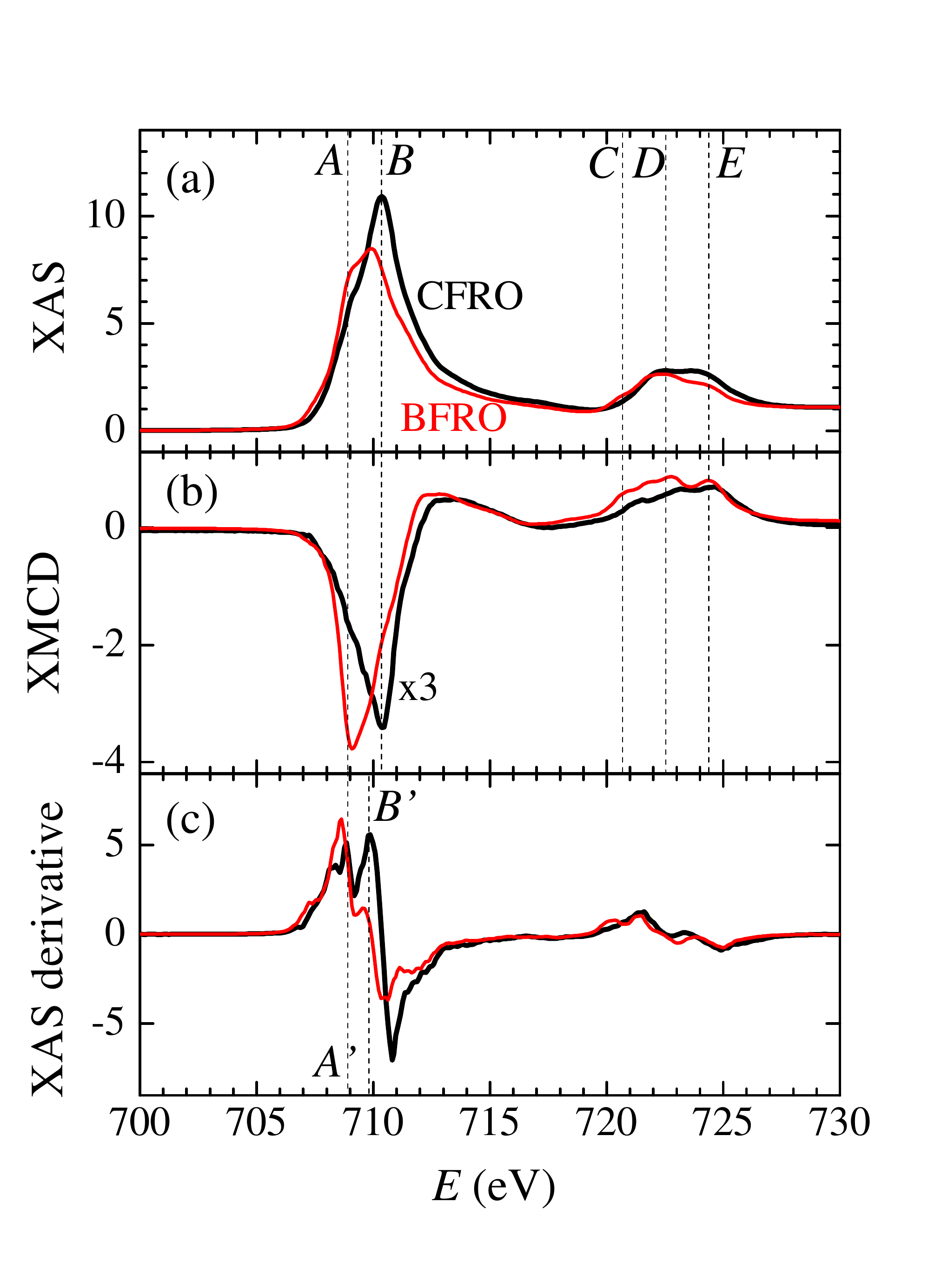}
\caption{\label{FeL23} (color online) (a) X-ray absorption (XAS) and (b) x-ray magnetic circular dichroism (XMCD), and (c) XAS derivative spectra of Ca$_2$FeReO$_6$ (Ca$_2$FeReO$_6$) and Ba$_2$FeReO$_6$ (Ba$_2$FeReO$_6$) at the Fe $L_{2,3}$ edges, taken at $T=10$ K with an applied field of 1 T for Ca$_2$FeReO$_6$ and 5 T for Ba$_2$FeReO$_6$. The dashed lines signal the features $A-E$ and $A'$, $B'$ discussed in the text.}
\end{figure}

\begin{figure}
	\includegraphics[width=0.45\textwidth]{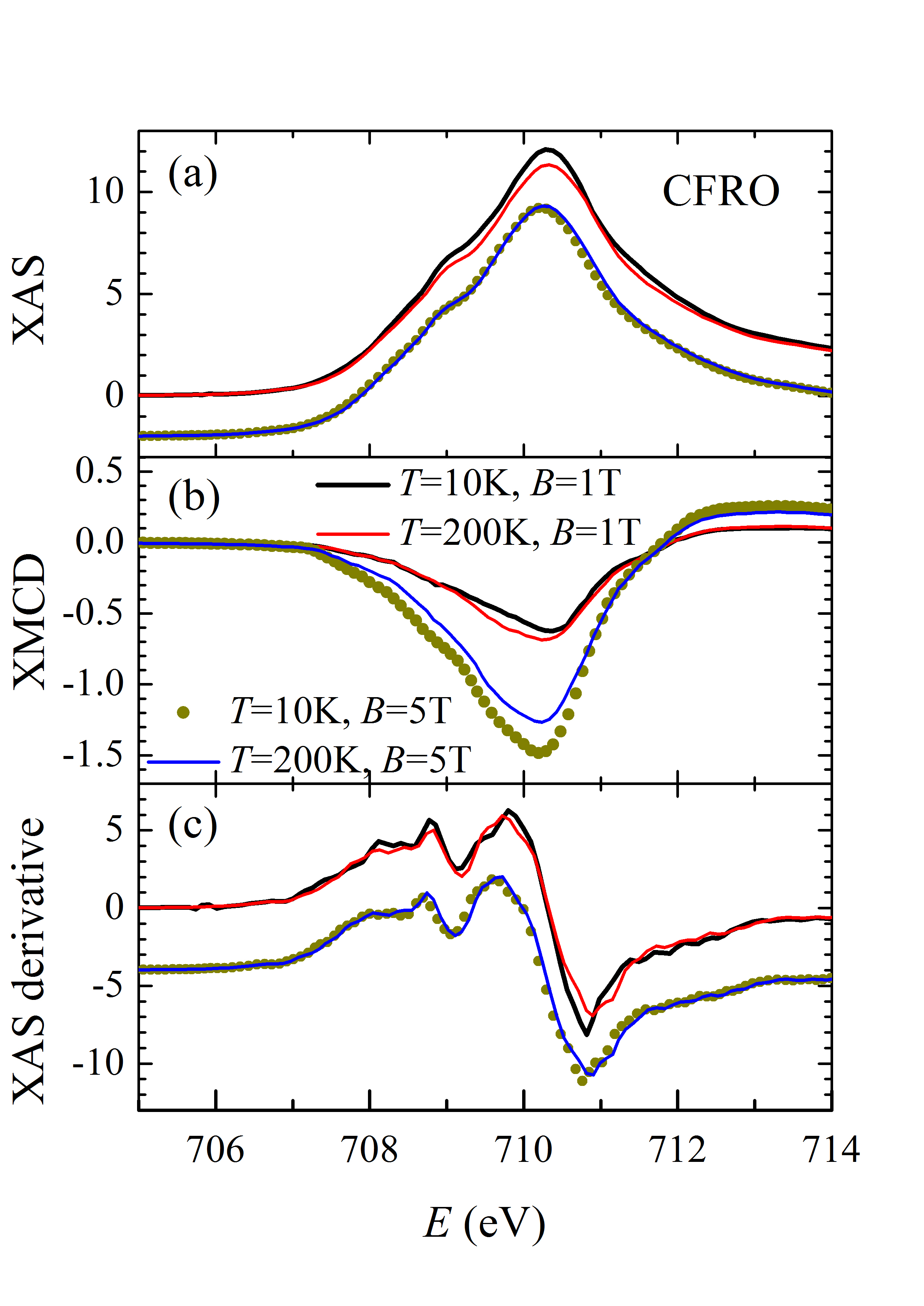}
	\caption{\label{CFRO_zoom} (color online) Fe $L_{3}$ edge x-ray absorption (XAS, a), x-ray magnetic circular dichroism (XMCD, b), and XAS derivative (c) spectra of Ca$_2$FeReO$_6$ at $T=10$ and 200 K, and $B=1$ and 5 T. In (a) and (c), the spectra for $B=5$ T were vertically translated for better visualization.}
\end{figure}

Figures \ref{FeL23}(a)-\ref{FeL23}(c) show the XAS, XMCD and XAS derivative spectra, respectively, of Ba$_2$FeReO$_6$ and Ca$_2$FeReO$_6$ at the Fe $L_{2,3}$ edges 
and $T=10$ K. These edges correspond to electronic transitions from the Fe $2p_{1/2}$ ($L_2$) and $2p_{3/2}$ ($L_3$) core levels to the Fe $3d$ states above the Fermi level. For the XAS spectrum of Ca$_2$FeReO$_6$ at the $L_3$ edge, a relatively sharp peak is observed at 710.5 eV (feature $B$ in Fig. \ref{FeL23}) with a shoulder at $\sim 709.0$ eV ($A$), while at the $L_2$ edge a structure with two overlapping peaks at 722.5 and 724.5 eV ($D$ and $E$) is seen. The XMCD spectrum of Ca$_2$FeReO$_6$ shows a negative peak at $B$ position in the $L_3$ edge and three positive peaks at the $L_2$ edge. 
The $L_3$ XMCD spectrum of Ba$_2$FeReO$_6$ shows a pronounced negative peak at the $A$ position, while the $L_2$ XMCD shows an additional positive peak at $C$ with respect to Ca$_2$FeReO$_6$. We should mention that a theoretical XAS spectrum of Ba$_2$FeReO$_6$ was recently generated by {\it ab}-initio calculations \onlinecite{Antonov}, showing very good agreement with our experimental data. The XAS derivative spectrum of Ca$_2$FeReO$_6$ shows two relatively sharp positive peaks at 708.8 and 709.8 eV, which are marked in Fig. \ref{FeL23}(c) as $A'$ and $B'$, respectively. Also, a shoulder located $\sim 0.7$ eV below feature $A'$ is observed. Measurements performed in different beamtime periods and on different pieces of our ceramic sample of Ca$_2$FeReO$_6$ showed variations of the relative magnitude of this shoulder with respect to the sharp feature $A'$, indicating a large sensitivity of this spectral feature on possible variations of the conditions of the probed surface. The systematic XAS and XMCD measurements of Ca$_2$FeReO$_6$ as a function of temperature and magnetic field shown below were taken in a fresh re-scraped ceramic piece yielding the most pronounced sharp $A'$ feature with a minimal low-energy shoulder as displayed in Figs. \ref{FeL23}(c) and \ref{CFRO_zoom}(c).


\begin{figure}
\includegraphics[width=0.5\textwidth]{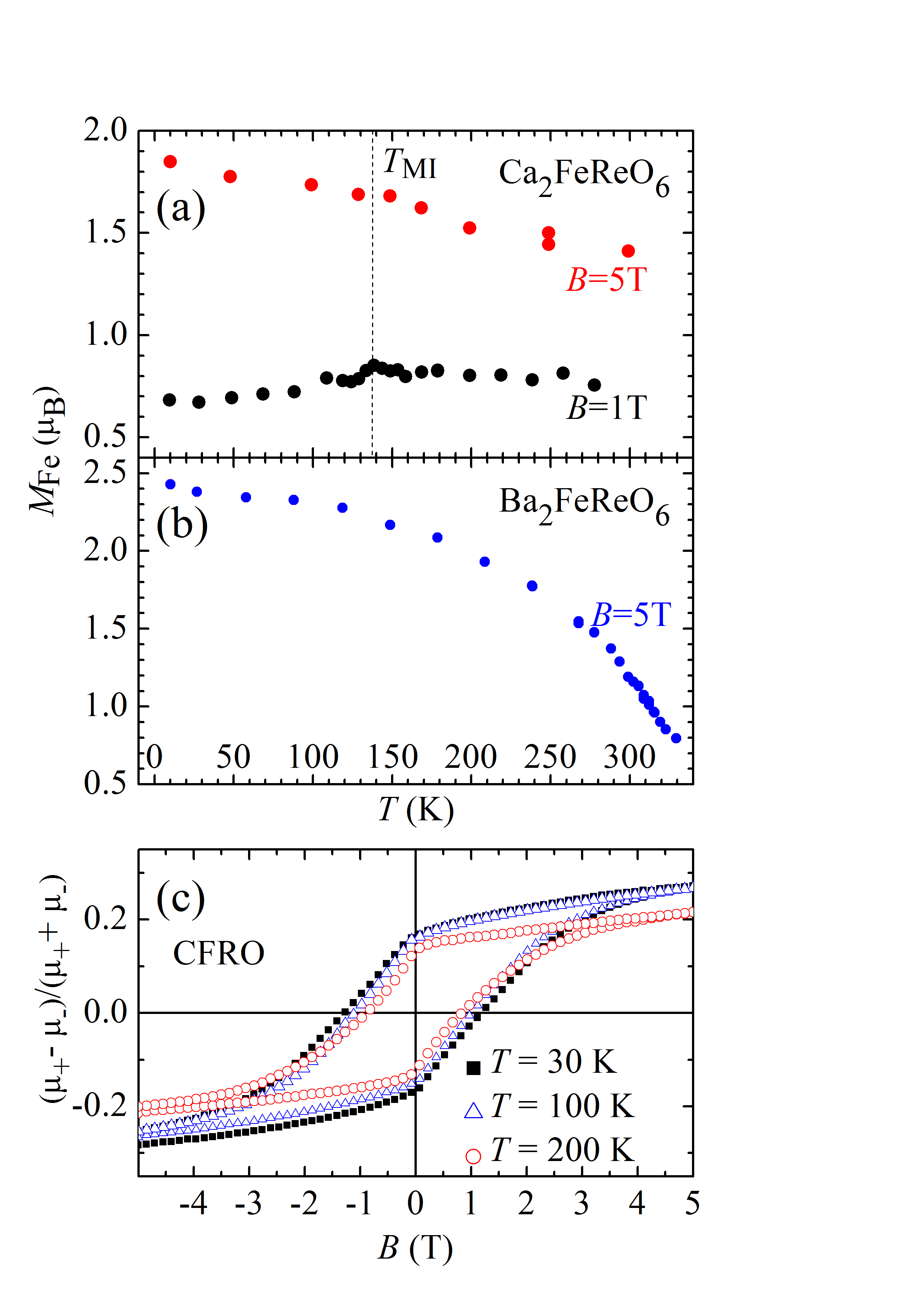}
\caption{\label{FeTdepMAG} (color online) Temperature dependence of Fe spin moments extracted through the XMCD sum rules at the Fe $L_{2,3}$ edges for Ca$_2$FeReO$_6$ at $B=1$ and 5 T (a) and for Ba$_2$FeReO$_6$ at $B=5$ T (b). (c) Magnetic field hysteresis loop of the Fe $L_3$ edge XMCD signal for Ca$_2$FeReO$_6$ at fixed $E=710$ eV and selected temperatures. Statistical error bars are smaller than the symbol size.}
\end{figure}

Figures \ref{CFRO_zoom}(a)-\ref{CFRO_zoom}(c) show the XAS, XMCD and XAS derivative spectra, respectively, of Ca$_2$FeReO$_6$ at the Fe $L_{3}$ edge at selected temperatures ($T=10$ and 200 K) and magnetic fields ($B=1$ and 5 T). Although all spectra are very similar, minor changes in the spectrum at $T=10$ K and $B=1$ T can be noticed, most notably sharper XAS derivative features at 709.5 and 710.5 eV [see Fig. \ref{CFRO_zoom}(c)]. These distinctions are most likely associated with the dominance of the insulating state below $T_{MI}$ and low fields, which may show slightly different lattice parameters  \onlinecite{Granado,Oikawa} and electronic structure, thereby altering slightly the Fe $L_{3}$ edge spectrum. For high fields ($B=5$ T) and/or high temperatures ($T=200$ K), the metallic phase tends to be dominant \onlinecite{Granado}, explaining the similar Fe $L_{2,3}$ XAS and XMCD spectral shapes for $B=5$ T obtained at both 10 and 200 K (see Fig. \ref{CFRO_zoom}).

From the XMCD data at the Fe $L_{2,3}$ edges, the powder-averaged Fe spin and orbital net moments aligned along the field direction were extracted through well established sum rules \onlinecite{Thole,Carra,Chen,Azimonte1,Azimonte2}. Here, we employ the Fe $3d$ level occupations $n_{3d}=5.84$ and $n_{3d}=5.93$ for Ca$_2$FeReO$_6$ and Ba$_2$FeReO$_6$, respectively, obtained by band structure calculations \onlinecite{Wu}, and also applied
the correction term $1/0.685$ and $2/(0.875+0.685)$ for the Fe spin moments in Ca$_2$FeReO$_6$ and Ba$_2$FeReO$_6$, respectively, due to the spin-orbit coupling in the Fe $2p$ core holes \onlinecite{Teramura,Azimonte1,Azimonte2}. Also, the mean value of the dipole magnetic operator $<T_z>$ is assumed to be much smaller than the projected spin $<S_z>$. This assumption is certainly valid for nearly spherical shells, which is arguably the case for Ca$_2$FeReO$_6$ considering the proximity of the Fe electronic configuration to the Fe$^{3+}$ state for Ca$_2$FeReO$_6$ (see below). For Ba$_2$FeReO$_6$ with an intermediate Fe valence between 2+ and 3+, this assumption is less justified. Nonetheless, even in this case we argue that the $<T_z>$ correction is rather modest, since an estimated $\sim 90$ \%\ of the Fe magnetic moments would arise from the fully occupied spin-up $3d$ band for which $<T_z>=0$. The extracted Fe moments at $T=10$ K and $B=5$ T are $M_{Fe}=1.85(15)$ and $2.43(15)$ $\mu_B$ for Ca$_2$FeReO$_6$ and Ba$_2$FeReO$_6$, respectively, which are consistent with previous results \onlinecite{Azimonte1,Azimonte2}. Figure \ref{FeTdepMAG}(a) shows the temperature-dependence of the Fe spin moments for Ca$_2$FeReO$_6$ for $B=1$ and 5 T. The corresponding data for Ba$_2$FeReO$_6$ and $B=5$ T are given in Fig. \ref{FeTdepMAG}(b). The Fe orbital moments are null within our sensitivity ($0.15$ $\mu_B$) for both samples and all temperatures and are not shown. The Fe spin moments for Ca$_2$FeReO$_6$ and $B=1$ T show a smooth increase on warming from the lowest temperatures, peaking at $T_{MI}$ and decreasing again on further warming. The thermal evolution of the Fe spins is qualitatively different for the same sample and $B=5$ T, rather showing a continuous decrease on warming with a small bump at $T_{MI}$. For Ba$_2$FeReO$_6$ and $B=5$ T, the Fe magnetization shows a continuous decrease upon warming, with an inflection point at $T_C^{Ba}=305$ K. The significant Fe moments above this temperature are ascribed to a $B$-induced magnetic polarization in the paramagnetic phase.


\section{Discussion}

\subsection{Electronic structure}

The XAS spectra at the Fe $L_{2,3}$ edges [see Fig. \ref{FeL23}] provide important information on the Fe electronic states of Ca$_2$FeReO$_6$ and Ba$_2$FeReO$_6$. A comparison of our data with published results for Sr$_2$FeMoO$_6$ and related compounds \onlinecite{Ray,AbbateDP,Kang,Kuepper,Kuepper2,Besse,Kang2} reveals that the $A$ shoulder is weaker for Ca$_2$FeReO$_6$ than for Sr$_2$FeMoO$_6$, indicating a valence state closer to Fe$^{3+}$ for the former. Also, the $L_3$ XAS spectrum of Ca$_2$FeReO$_6$ is similar to that of LaFeO$_3$ \onlinecite{Abbate}, except for a smaller splitting of $A$ and $B$ peaks and a slightly larger $A/B$ spectral weight ratio for Ca$_2$FeReO$_6$. These considerations indicate an Fe valence state close to +3 for our Ca$_2$FeReO$_6$ sample, which is consistent with a M\"osbauer spectrum obtained for this sample \onlinecite{Gopal} and with bond valence analysis using the Fe-O distances obtained from powder diffraction data, which yielded a valence of +2.94 at room temperature \onlinecite{Granado,Oikawa}. The smaller splitting of the $A$ and $B$ features in Ca$_2$FeReO$_6$ with respect to LaFeO$_3$ is associated with a smaller overall cubic crystal field parameter $10Dq$ for the former, which is presumably an influence of the strongly charged Re$^{5+}$ cations in the crystal field sensed by Fe. For Ba$_2$FeReO$_6$, the shoulder $A$ at the XAS $L_3$ edge shows a significantly larger spectral weight in comparison to Ca$_2$FeReO$_6$. Also, at the $L_2$ edge an additional peak component ($C$) can be seen at 719 eV. Note that, besides the changes in lineshape, the XAS and XMCD spectra of Ba$_2$FeReO$_6$ are shifted to lower energies in comparison to Ca$_2$FeReO$_6$, indicating a smaller Fe valence state for the former. This conclusion is also in general agreement with previous XAS \onlinecite{Herrero}, M\"ossbauer  \onlinecite{Gopal}, and neutron diffraction \onlinecite{Azimonte1} studies.

It is interesting to notice that, according to the above scenario, the formal Re valence must be close to Re$^{5+}$ in the insulating phase of Ca$_2$FeReO$_6$, with two electrons in the Re $5d$ levels. In an atomistic picture, this should lead to local  states with effective total angular momentum $j=2$ \onlinecite{Chen}. The local Re magnetization in this case is then given by ${\bf M}$(Re)$={\bf j}/2$ in units of Bohr magnetons \onlinecite{Chen2}, thus the ordered local magnetization projected into the principal axis should be $M_y$(Re) $=1$ $\mu_B$, in agreement with experimental values obtained by neutron diffraction \onlinecite{Granado,Oikawa}.


\subsection{Magnetism}

A detailed account of our XMCD measurements provides useful information on the element-specific magnetism of Ca$_2$FeReO$_6$ and Ba$_2$FeReO$_6$. As shown in Figs. \ref{FeTdepMAG}(a) and \ref{FeTdepMAG}(b), the extracted Fe moments at $T=10$ K and $B=5$ T through XMCD sum rules are $M_{Fe}=1.85(15)$ and $2.43(15)$ $\mu_B$ for Ca$_2$FeReO$_6$ and Ba$_2$FeReO$_6$, respectively, which should be contrasted with the zero-field neutron diffraction values at comparable temperatures, $M_{Fe}=3.42(7)$ and $3.16(10)$ $\mu_B$, respectively \onlinecite{Granado,Azimonte1}. Thus, while the XMCD moment taken  for Ba$_2$FeReO$_6$ at $B=5$ T is 77 \% of the neutron diffraction value, for Ca$_2$FeReO$_6$ this proportion is reduced to only 54 \%. In a previous preliminary study, such surprisingly small moments were attributed to magnetically hard magnetic domains and significant Fe/Re intersite disorder across those grain boundaries probed by soft XAS measurements obtained by total electron yield detection method \onlinecite{Azimonte2}. On the other hand, the Re moments previously obtained by bulk-sensitive XAS at the hard x-ray Re $L_{2,3}$ edges at ordinary fields ($B \leq 5$ T) also seem to be significantly reduced with respect to neutron diffraction values \onlinecite{Escanhoela,Azimonte1,Granado,Sikora1}. Also, the $M \times B$ curves of Fig. \ref{FeTdepMAG}(c) indicate that the Ca$_2$FeReO$_6$ moments are close to domain wall saturation for $B=5$ T in the conditions of our experiment, and therefore the relatively small Fe spin moments aligned to such an external field obtained by XMCD do not seem to be attributable to persistent magnetic domains. Note that the extremely large magnetocrystalline anisotropy of the Fe moments implied by the above considerations is highly unusual for Fe$^{3+}$ with half-filled $3d$ shell and null orbital moment, confirmed here by our analysis of XMCD data. In the present case, the hardness of the Fe spins is presumably caused by a superexchange interaction between Fe $3d$ and Re $5d$ moments, where the latter can be pinned to the lattice by the large unquenched orbital component. For Ca$_2$FeReO$_6$, the monoclinic structure likely stabilizes an specific spin-orbital configuration for the Re $5d$ moments, also pinning down the Fe spins. As a consequence, the Fe moments would be necessarily confined to a specific axis or plane of the crystal structure. In a powder ceramic sample like the present case, the crystalline axes are randomly oriented with respect to the applied magnetic field, therefore the component of the average Fe moments along the field measured by XMCD should be substantially smaller than the spontaneous Fe moments measured by neutron diffraction, as indeed observed here. A simple calculation shows that, when the atomic moments are constrained to lie along an specific axis of the crystal structure, the domain-saturated powder-averaged magnetization along the field direction will be only $M/2$, where $M$ is the spontaneous intradomain magnetization along the easy axis (probed by neutron scattering). If the above constraint is partly relaxed so that the moments have an easy plane rather than an easy axis, the saturated powder-averaged magnetization along the field direction will be $\pi M/4=0.78M$. The XMCD value for the Fe moment for $B=5$ T is 54 \%\ of the spontaneous moment obtained from neutron powder diffraction for Ca$_2$FeReO$_6$ at low temperatures, which is close to the expected value for an easy axis configuration. We should mention that, for much higher fields ($B \sim 30$ T), the magnetic moments will tend to align along the field direction through a non-hysteretic magnetic rotation process, leading to magnetic moments obtained by XMCD that should be closer to the spontaneous moments obtained by neutron diffraction \onlinecite{Sikora2}. 

The above considerations provide insight not only to the harder magnetic state of the insulating phase with respect to the metallic one, but also to the physical mechanism leading to the magnetic field-dependence of the balance between these phases \onlinecite{Granado}. In fact, since the metallic phase has a easy-plane spin-orbital configuration and therefore a larger powder-averaged magnetization in the field direction with respect the easy-axis configuration of the insulating state, that phase also has a stronger Zeeman coupling energy, thereby altering the delicate energy balance between such competing phases. The influence of the magnetic field is amplified by the fact that the competing phases have very similar free energies over a wide temperature range \onlinecite{magnetoresistance}.

For the quasi-cubic tetragonal crystal structure of Ba$_2$FeReO$_6$, the situation may be distinct with respect to Ca$_2$FeReO$_6$, owing to the quasi-degeneracy between the Re spin-orbital states. In this case, the magnetic moments may be allowed to align along equivalent crystallographic directions, such as the {\bf a}, {\bf b}, or {\bf c} quasi-cubic axes, whichever is closest to the external field for each grain. This scenario tends to yield an average projected Fe moment in the field direction, measured by XMCD at relatively modest fields, which is only slightly smaller than the neutron diffraction value, consistent with our observation for this material. A weaker magnetostructural coupling for this compound is also consistent with the rather weak tetragonal distortion below $T_C$ \onlinecite{Azimonte1,Ferreira} and the nearly gapless magnetic excitation spectrum \onlinecite{Plumb}.

\section{Conclusions}

In summary, the Fe electronic state in $A_2$FeReO$_6$ ($A$ = Ca and Ba) is found to be strongly dependent on $A$. For Ca$_2$FeReO$_6$ the Fe valence state is close to 3+ both below and above $T_{MI}$. For Ba$_2$FeReO$_6$, the Fe $L_{2,3}$ XAS spectrum is consistent with an intermediate state between Fe$^{2+}$ and Fe$^{3+}$. Such valence instability of Fe with respect to an isoelectronic $A$-site substitution likely results from a competition between (i) Re $5d$ electronic correlations assisted by strong spin-orbit coupling favoring an insulating ground state with pure Fe$^{3+}$ and Re$^{5+}$ valence states, and (ii) double exchange interactions for hybridized Fe($3d$)-O($2p$)-Re($5d$) electrons that favor a ferrimagnetic and metallic ground state with mixed-valent Fe and Re ions. Concerning the magnetic properties, the powder-averaged Fe moments aligned to a field of 5 T obtained through XMCD sum rules for Ca$_2$FeReO$_6$ is nearly half of the spontaneous moments extracted from previous neutron diffraction data at low temperatures, suggesting a very strong magnetocrystalline anisotropy for the Fe $3d$ local moments in this material, presumably caused by exchange interactions with the Re $5d$ moments. This effect is less severe for Ba$_2$FeReO$_6$.

\begin{acknowledgements}

ESRF is acknowledged for concession of beamtime. This work was supported by FAPESP Grants No. 2017/10581-1 and No. 2018/20142-8, and CNPq Grants No. 409504/2018-1 and No. 308607/2018-0, Brazil.

\end{acknowledgements}

\end{document}